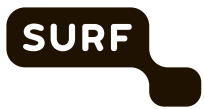

# Whitepaper

# Deep-learning enhancement of large scale numerical simulations

**March 2020**

**Caspar van Leeuwen, Damian Podareanu, Valeriu Codreanu, Maxwell Cai, Axel Berg**
SURF Open Innovation Lab, SURF

**Maxwell Cai, Simon Portegies Zwart**
Leiden Observatory, Leiden University

**Robin Stoffer, Menno Veerman, Chiel van Heerwaarden**
Meteorology and Air Quality Group, Wageningen University and Research

**Sydney Otten, Sascha Caron**
Institute for Mathematics, Astro- and Particle Physics IMAPP, Radboud University

**Cunliang Geng, Francesco Ambrosetti, Alexandre M.J.J. Bonvin**
Bijvoet Centre for Biomolecular Research, Faculty of Science - Chemistry, Utrecht University

**Contact:**
high_performance_machine_learning@surfsara.nl

## Summary

Traditional simulations on High Performance Computing (HPC) systems typically involve modelling very large domains and/or very complex equations. HPC systems allow running large models, but limits in performance increase that have become more prominent in the last 5-10 years will likely be experienced. Therefore new approaches are needed to increase application performance. Deep learning appears to be a promising way to achieve this. Recently deep learning has been employed to enhance solving problems that traditionally are solved with large-scale numerical simulations using HPC. This type of application, deep learning for high performance computing, is the theme of this whitepaper. Our goal is to provide concrete guidelines to scientists and others that would like to explore opportunities of applying deep learning approaches in their own large-scale numerical simulations. These guidelines have been extracted from a number of experiments that have been undertaken in various scientific domains over the last two years, and which are described in more detail in the Appendix. Additionally, we share the most important lessons that we have learned.

# Contents

# Introduction

Traditional numerical simulations on High Performance Computing (HPC) systems typically involve modelling very large domains and/or very complex equations, such as computational chemistry, computational fluid dynamics, material sciences, astrophysics and high energy physics. HPC systems allow running large models, but users will experience limits in performance increase that have become more prominent in the last 5-10 years.

One of them is that Moore's law is slowly coming to an end, caused by the exponential growth in costs of microprocessors with reduced chip component size reaching nearly physical limits. Another one is Amdahl's argument where maximum speedup levels are reached by the ever-increasing required levels of parallelism of large HPC systems. This means that, unless future transistors will be designed in a fundamentally different way, it is doubtful whether the continuous increase in computational power that we have become used to in the past decades will continue. *New approaches* are needed in order to allow application performance increase up to the level that we are used to. So, how do we enable larger and more complex models in the future?

Over the past few years, deep learning has received a lot of attention. Deep learning has proven to be a very versatile method, capable of solving classification problems [1], generation problems [2], speech recognition [3] [4], and many more. More recently, deep learning was employed to enhance solving problems traditionally solved with large-scale HPC simulations [5] [6]. This type of application we refer to as *deep learning for high performance computing* (or *deep learning for HPC* for short). These deep learning models have the potential to reach higher levels of accuracy than the traditional models and/or reduce the time to solution.

To explore what deep learning can do to improve traditional numerical HPC simulations, SURF collaborated with four Dutch university groups from various scientific domains. The goal was to learn how deep learning could best be leveraged (e.g. would it be a pre- or post-processing step? Or would it replace simulation kernels altogether?) as well as which types of simulations are most suitable to enhance or accelerate using deep learning.

This white paper presents the lessons learned from these projects. First, we start with general guidelines for running a project that focuses on accelerating traditional HPC applications using deep learning. Then we report on the individual five projects we have run (one collaboration resulted in two projects). The goal is to provide guidelines to scientists as well as others that want to start similar projects in their own scientific discipline.

Throughout this document, we will assume some basic familiarity with neural networks, and neural network terminology. If you are not familiar with these networks, we advise you to consult one of the many basic introductions on the web. Whenever a reference is being made to one of the use case projects that are described in the Appendix, the title of the respective project is written in italics.

# Overall approach and steps to be taken

In this paragraph we sketch an overview of our overall approach and the steps to be taken when you are thinking about employing deep learning for HPC (Figure 1). In the paragraph's hereafter we describe each of the steps in more details.

The first step would be to understand your simulation, to find out where you might profit most from deep learning. You might for instance use a profiling tool to find out which routines take the most time. Or you might reduce the number of individual simulations by investigating how to explore the parameter space in a more clever way.

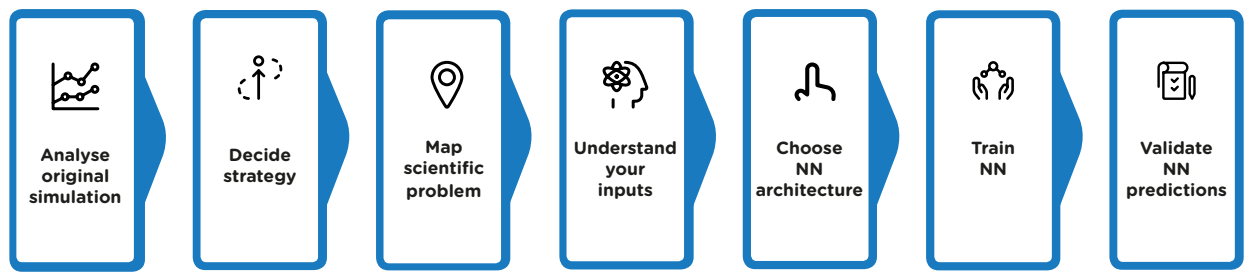

**Figure 1. Steps to take when augmenting a traditional HPC simulation with deep learning.**

Once you understand your simulation, you have to decide on your strategy. Two basic approaches can be distinguished: augmenting a simulation, or replacing (parts of) a simulation by deep neural networks. In both cases, you can choose whether to focus on reducing the computational effort or increase the accuracy. The more fundamental approach is replacing (part of) a simulation with a deep learning model. In doing so, you are also replacing an analytical model with an empirical model. You can either replace the full simulation, or replace a kernel, typically a part that is expensive to compute.

If you are clear about your strategy, the next step is to map your scientific problem. Often, there are many ways to look at the same problem. For example, a meteorological simulation of the atmosphere could be approached as a pattern recognition problem or as a time series problem. In deciding on the most effective approach, you should ask yourself three questions: (1) Will my chosen mapping create a scenario where the relationship between the input and output space can be learned? (2) Do I have enough data to train on if I use that mapping?, and (3) Is that data of sufficient quality?

The mapping also has implications for your choice of neural network: the NN architecture. A 'natural' mapping between your problem and the chosen architecture is important. The type of data is the prime issue here. For instance, a fully connected architecture is suitable for flat unstructured data, whereas with 2D of 3D spatial data a convolutional architecture is preferable. Another consideration in choosing the architecture is the task it needs to perform. For instance, if you have a problem that is more of a stochastic nature (one that would return an output that is a sample from some probability distribution), you may need to look into generative network architectures.

Once you have chosen a way to map your problem, you'll need to think about how to encode your inputs. Neural networks may have difficulty if your inputs have large dynamical ranges, include rare events or irrelevant information. This last issue can be solved by normalizing your data: this reduces dynamical ranges, but can also remove irrelevant information from your input. If your dataset is unbalanced (if some types of samples are overrepresented), you have to balance it, for instance by resampling the rare events more often.

The next step is designing a neural network, preferably a model that works well for your type of problem. In some cases, several predefined architectures are available. However, you often will need to tune your model. This requires both expertise in neural network engineering and an intricate understanding of the system that is being modelled. It is a task that requires intensive collaboration between an AI expert and an expert of the scientific simulation being modelled.

The final step is the validation of the model. The easiest way is to compare the predictions against those made using the traditional simulation. When you have replaced a kernel, validation should be done in two stages. Firstly, you benchmark your neural network predictions against the outputs of that (traditional) kernel specifically (a priori validation). Secondly, the neural network should be integrated into the traditional simulation, replacing the kernel it was trained on. Then, the results of this simulation should be benchmarked against the traditional simulation. This is called an a posteriori validation.

## Understanding your traditional numerical simulation

In any traditional form of code optimization, you need to understand your simulation. What kind of routines are executed as part of the simulation, in which order and what are their dependencies? And, probably most importantly: which routines take the most time? In order to answer that question, a profiling tool is typically used to analyze your program (Figure **2**). The output of such a profiling tool will tell you which part of the program should be focused on to reduce the overall runtime of your program as much as possible.

Another issue is that some individual simulations may not be computationally heavy at all, but typical usage would involve running *many* of them. For example, one may need to run a parameter sweep on one of the simulation's parameters, or run an ensemble simulation to make predictions for an intrinsically chaotic system. In that case, it may be more worthwhile to investigate how to explore the parameter space in a cleverer way, rather than speeding up the individual simulation.

In a *deep learning for HPC* approach, a similar understanding of your traditional numerical simulation code and its use is essential.

**Whitepaper** | Deep-learning enhancement of large-scale numerical simulations

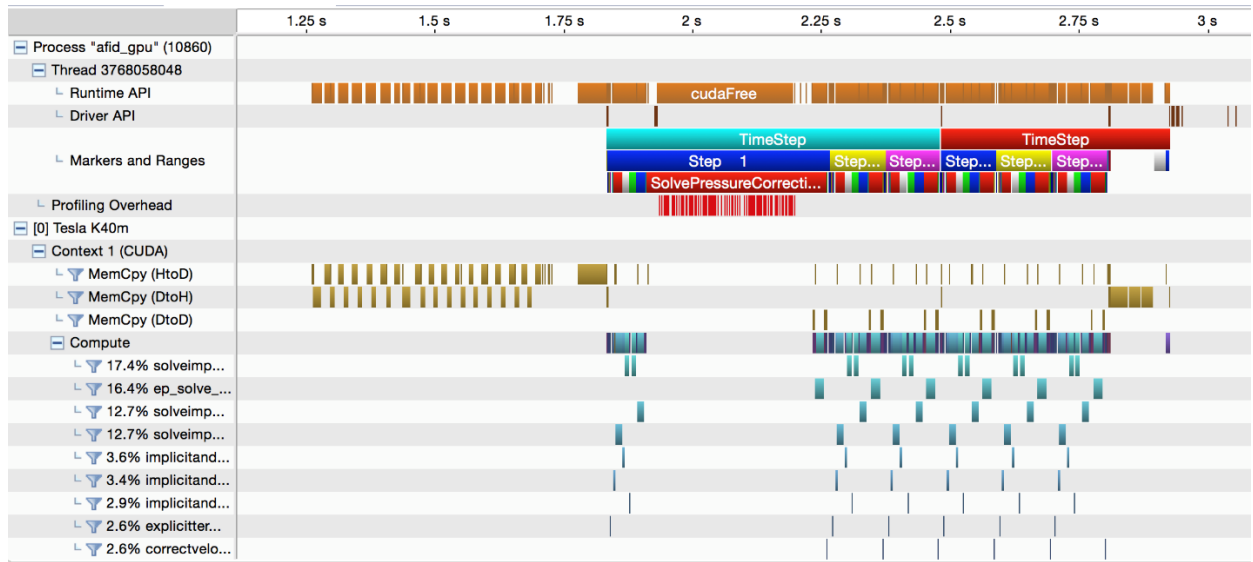

**Figure 2: Example output of the NVIDIA Visual Profiler.**

# Strategies to enhance or speed up an HPC simulation with deep learning

Two basic approaches can be distinguished: augmenting a simulation, or replacing (parts of) a simulation by deep neural networks. In augmenting the simulation, one doesn't directly replace one of the simulation kernels, but rather adjust another part such as the pre- or post-processing step in order to reduce the amount of computation needed. Some examples are:

- Instead of performing a naïve grid search to scan a parameter space, use a machine learning or deep learning model to predict which points in the parameter space are the most important for you to evaluate.
- Instead of randomly initializing an iterative algorithm, use a deep learning model to predict a 'good' initialization.
- Run a simulation at relatively low resolution, and use a deep learning algorithm to refine the output. For example, Google has developed a method to increase the resolution of images using a neural network [7].
- Similar to the previous point, but specific to a grid search: run a coarse grid search and use deep learning to interpolate the results.

In each of these approaches, one can usually choose whether to focus on reducing the computational effort (e.g. by scanning fewer points in a parameter space) or increase the accuracy (e.g. keep the number of points scanned in a parameter space the same).

The second and more fundamental approach is to replace (part of) a simulation with a deep learning model. Traditional HPC simulations are generally based on analytical models which map some input space to an output space. For example, a simulation could be mapping an initial state (e.g. today's atmosphere) to a final state (tomorrow's atmosphere). Another option would be to map some parameters (e.g. a molecule in a certain conformation) to an output (the energy for that conformation). The same is generally true for kernels that compose a simulation. Deep learning models are also designed to map an input space to an output space, but in an empirical fashion.

The key difference between analytical and empirical models is that analytical models are based on fundamental scientific theory, while empirical models are derived from data. There are a few key differences between the two:

| Analytical models | Empirical models |
| --- | --- |
| Driven by scientific theory | Driven by data |
| Explainable | Limited explainability |
| (Often) based on assumptions | No/few assumptions |
| Subject to the quality of the assumptions | Subject to the quality and quantity of the data |
| Making predictions can be expensive | Making predictions is (generally) cheap |

In line with the differences mentioned above, some challenges and opportunities may be expected when trying to replace or augment traditional HPC simulations with deep learning:

- Since deep learning is not driven by theory and since it is often more difficult to explain deep learning models, acceptance of such models in a scientific community may be challenging
- Deep learning models depend on the quality and quantity of the data. Training them using (analytically) simulated data is often more feasible than using measured data, but it should be realized that the training results will be limited by the accuracy of the simulation they were based upon. Thus, deep learning enhanced HPC provides clear opportunities in scientific domains where very accurate models are available, but where approximative models are used in practice for computational efficiency. The accurate models can then be used in a one-time effort to train a deep learning model, after which the deep learning model may be able to provide both accurate and computationally cheap results.
- Related to the previous point: experts sometimes have to make pretty strong assumptions (even when they know these may not be completely satisfied) to design simulations that are computationally cheap enough. Deep learning models have the advantage that the model may learn such assumptions (if they are true), but they don't have to be put in explicitly. This reflects a more generic advantage of a machine learning approaches over expert designed systems, namely that they learn what is important, rather than an expert having to make that decision explicitly.

When replacing (part of) a simulation with deep learning, two options are available (see Figure 3):

- Replace the full simulation, e.g. use a neural network to predict tomorrow's weather using the current state of the atmosphere as input.
- Replace a kernel in a simulation, typically focusing on the parts that are most expensive to compute. For example: in multiscale modelling, one could replace only the kernels that simulate the finer scale. Another example: to model radiation transport in the atmosphere, the radiative properties in the grid are first computed based on chemical composition, pressure, etc. Computing the radiative properties could be done using a neural network.

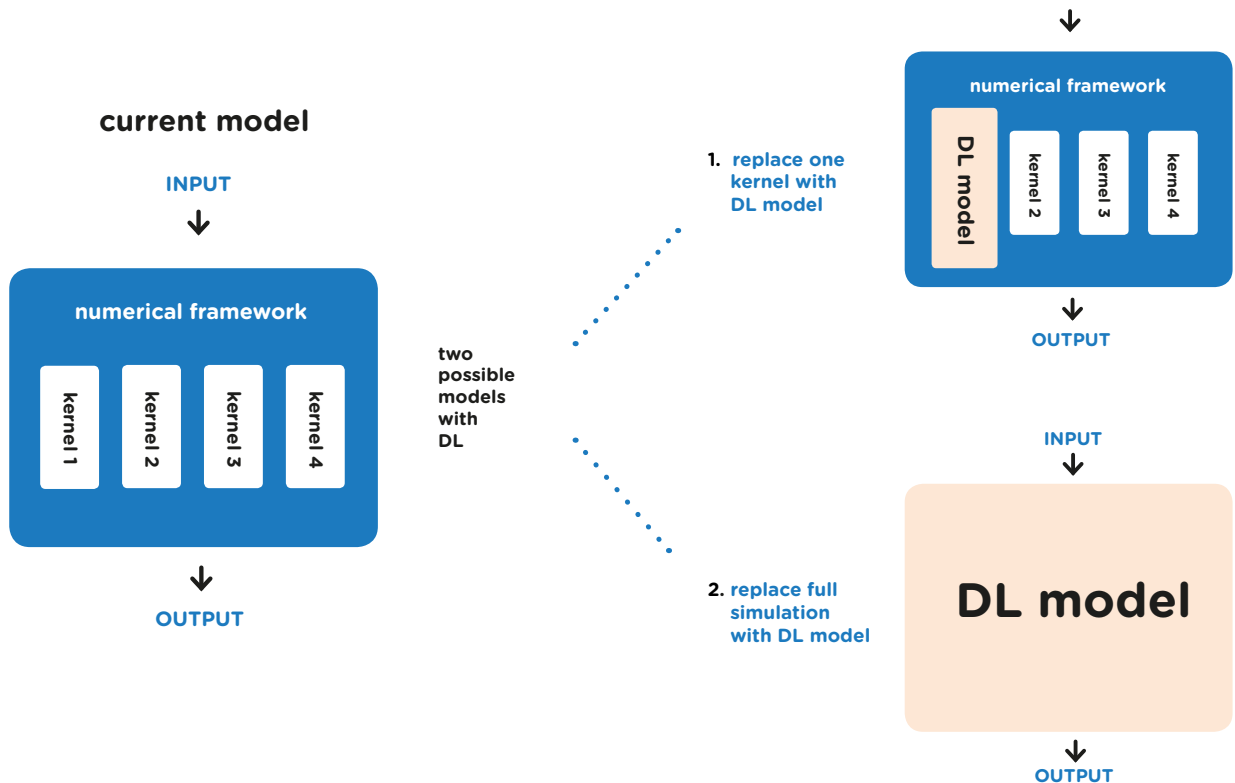

**Figure 3. Two strategies in replacing (part of) a simulation by a neural network.**

Just like in traditional optimization, the speedup from replacing a subroutine will always be limited: if the subroutine took 50% of the overall runtime of your original code, you'll never be able to accelerate your code with more than a factor of 2. It is one of the reasons why profiling is important: in general, you'll want to focus on a subroutine that accounts for a substantial amount of your total runtime. The potential speedup from replacing the entire code could be enormous (our project *Generating physics events without an event generator* followed this strategy). However, it may be easier to create a neural network that can successfully replace your subroutine, and it may also be easier to have such a replacement accepted by your scientific community (our project *Machine-learned turbulence in next-generation weather models* followed this strategy).

## Mapping your scientific problem

Once you have selected a strategy to enhance your simulation, the next thing to consider is how to map your scientific problem. Often, there are many ways to look at the same problem. For example, Figure **4** a (meteorological) simulation of the atmosphere could be approached as a pattern recognition problem (because it relates to 3D spatial data), or as a time series problem (because e.g. flow velocity in a single voxel is a time series). Approaching a meteorological simulation as a spatial problem, we might decide to try and predict the condition of tomorrow's atmosphere based on that of today. This would mean taking a 3D space as input, and trying to produce a 3D space as output – a problem very close to traditional image-to-image translation tasks [8]. Approaching it as a temporal problem, we might look at the atmospheric conditions for each voxel individually, interpret them as a time series, and use a sequence analysis method (e.g. recurrent neural networks) to predict the next time point (tomorrow's atmosphere).

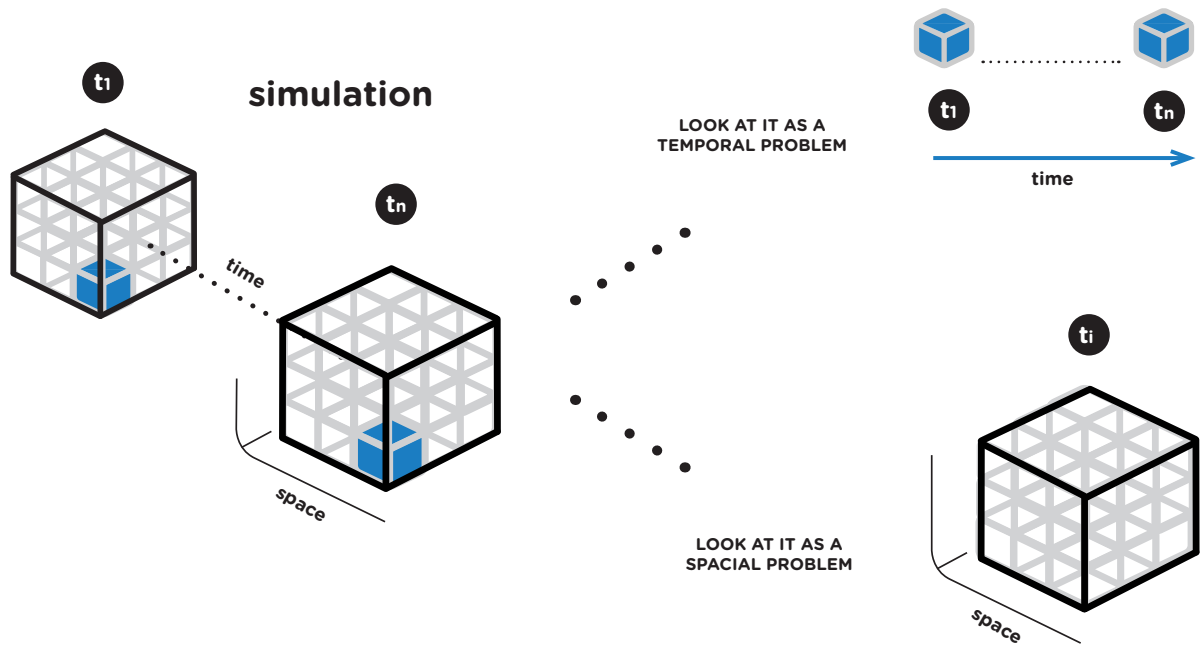

**Figure 4. A simulation of e.g. the atmosphere describes the state of the atmosphere as a 3D grid, throughout time. From a deep learning perspective, this can be interpreted as a temporal or spatial prediction problem.**

There are a number of key questions to be answered when deciding on a suitable mapping:
- Will my mapping create a scenario that allows me to learn about the relationship between the input and output space?
- Do I have enough data to train on if I use that mapping?
- Is that data of sufficient quality?

The first question is a very generic question that applies to all deep learning tasks, regardless of whether they have a scientific goal or not. However, in this scenario, the scientific domain expert will probably be the only one able to answer this question.

An example of a 'learnable relationship' can be found in our meteorology project. Here, one of the goals was to have a neural network estimate turbulent motions smaller than the grid spacing of the traditional simulation, using this coarse simulation data as input. Naively, one might think that it is impossible to know anything about what is going on *within* a grid cell of a simulation. However, domain experts know

Whitepaper | Deep-learning enhancement of large-scale numerical simulations

that the turbulent motion at fine scales (i.e. smaller than the grid spacing) are related to the turbulent motion at coarse scales (i.e. the motions that *can* be resolved by the coarse simulation). Thus, the input *does* contain information about the desired output and thus there *is* a relationship here that can, potentially, be learned.

The fact that you know there is *some* relationship between input and output of course does not guarantee that the input contains sufficient information for the output to be predicted perfectly. In practice, you may have to experiment with different mappings to see in which cases your input *most completely* defines your output.

To illustrate how a chosen mapping can affect the amount and quality of data available, let us consider the atmospheric simulation example. Mapping this as a time series problem, a single simulation would provide a number of time series equal to the number of voxels (and thus a large number of data samples). However, the implicit assumption in this mapping is that the temporal relationship is independent of the spatial location of the original voxel. If that assumption does not hold, data quality can be considered poor. Mapping the same problem as a spatial problem, a single atmospheric simulation would provide a much smaller number of samples, but does not require the same implicit assumption. Thus, data quality is (potentially) higher.

So, which neural network architecture fits a certain mapping? A very rough indication is given in the table below

| Type of data | Architecture |
| --- | --- |
| Flat unstructured data | Fully connected |
| 2D / 3D spatial | Convolutional architecture |
| Time series | Recurrent neural network or 1D convolutions |

There are many variations to the architectures listed here, and picking the most suitable one is probably best done with the help of an AI expert. A 'natural' mapping between your problem and the chosen architecture is important. For example, a time series problem *could* be analyzed with a fully connected network, but because the architecture does not represent the problem in a natural way you may need a lot of degrees of freedom (and therefore a lot of training data). To compare with traditional function-fitting: you *can* fit a non-periodic function with a Fourier approximation, but you will need a large number of terms in the Fourier series to gain a reasonably accurate result and it will probably generalize poorly.

Another consideration in choosing the architecture is the task it needs to perform. If you need to predict a single number (or if you are interested only in some statistical average), you essentially have a regression problem: for each input, there is one output that is the 'right' answer. Sometimes, however, your problem is stochastic and your traditional simulation would return an output that is a sample from some probability distribution. In those cases, you may need to look into generative network architectures, such as the variational auto-encoder (also known as VAE) [9] or generative adversarial network (also known as GAN) [2]. Note that these generative architectures could still have e.g. convolutional layers, but they allow for modelling stochastic processes by sampling a randomly initialized vector from a latent space and using this as an input. Our project *Generating physics events without an event generator* is an example. Particle production in a particle collision event is a stochastic

Whitepaper | Deep-learning enhancement of large-scale numerical simulations

process that is traditionally simulated using Monte Carlo simulations which generate events based on known probability distributions. In order to replace such a simulation, a generative network was used. Note that with some additional inputs, a generator can for example be conditioned to generate certain types of events, e.g. only events that produce a certain particle type, etc.

## Understanding your inputs

Once you have chosen a way to map your problem, you'll need to think about how to encode your inputs. Neural networks may have difficulty if your inputs have:

- Large dynamical ranges
- Rare events
- Information that is *not* needed to predict the output

These are general issues in deep learning that can be solved with the aid of an AI expert.

It is fairly common to normalize input data. This reduces dynamical ranges, but can also remove irrelevant information from an input. For example, consider an image. The absolute brightness or contrast are often not meaningful in deciding whether the image contains a cat or a dog. Normalizing the image essentially removes variations in brightness and contrast in your dataset, thus allowing your neural network to focus on the things that really matter. Similar considerations apply to your scientific problem: unless the absolute scale or contrast of your inputs is meaningful in predicting your desired output, you probably want to normalize your inputs.

Another problem that deep learning struggles with is unbalanced datasets. That is: if some types of samples are overrepresented in a dataset, the neural network typically learns very well to predict for those samples, but poorly for samples that are underrepresented. There are many ways to deal with unbalanced datasets, the simplest of which is simply resample the rare samples more often. If you find that you've trained a neural network that works poorly on some particular type of inputs, consider if they are represent sufficiently in your dataset and if not explore methods to balance it.

Finally, a particularly interesting point for scientific simulations is the third one. Suppose that our analytical equations have invariances. There are three main ways to make sure the neural network reproduces these invariances:

- Enrich the input by applying these invariant transformations to it (known as dataset augmentation). As a simple example: a handwritten 'six' is still a six if a rotation is applied to it. Thus, to recognize handwritten digits, one could rotate all samples in the dataset over randomized angles and supply these to the neural network, so that the network *learns* this invariance (see Figure **5**).
- Preprocess your inputs so they display the desired invariance. E.g. suppose we want to make predictions based on a wind velocity field, but we know (from science) that the outputs are independent of wind direction. We can then remove the directional information from the inputs by taking the magnitude before we feed it to the neural network.
- Make sure your network incorporates the invariance intrinsically. E.g. convolutional layers are translation invariant. Thus, a network consisting exclusively of convolutional layers would also be translation invariant. Note that (fully) convolutional networks are somewhat rare, but it shows that certain types of layers represent these invariances better than others.

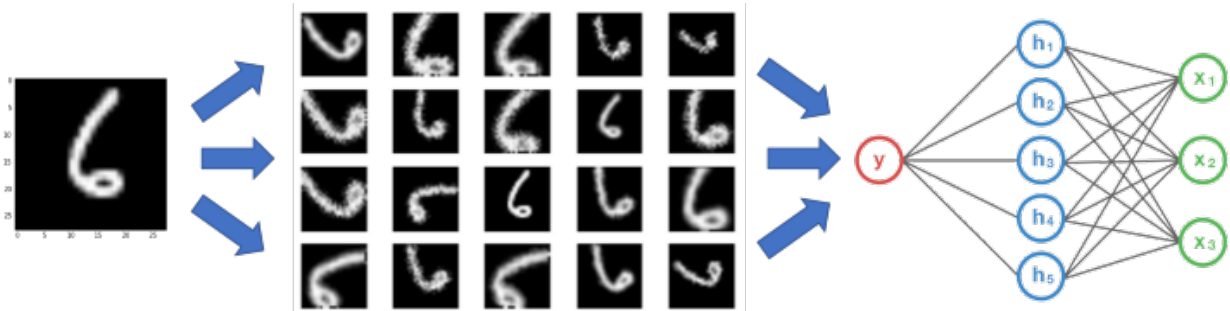

**Figure 5: dataset augmentation to allow a neural network to learn rotation invariance.**

## Designing a model

Designing a neural network architecture is an AI research task in itself. The most pragmatic approach is often to take a model that you know works well for *your* type of problem. For example, if you have an image-to-image translation task, there are several predefined architectures available (a good starting point might a neural network repository [10]).

After selecting and experimenting with an initial predefined architecture, your model may work well out of the box, but often you will have to tune it. The next step is then to find out *why* your network does not represent your problem well (which samples it fails to predict well for), and how it may be improved. This requires substantial expertise in neural network engineering, but also an intricate understanding of the system that is being modelled. As such, it is a task that requires intensive collaboration between an AI expert and an expert of the scientific simulation being modelled.

## Model validation

The easiest way to assess the quality of your deep learning model is compare it against predictions made using the traditional simulation. To test generalizability, it is important that these samples were not included in the dataset used to train the neural network.

In case you have replaced a full simulation, the comparison is rather trivial: simply compare the predictions of the neural network against those of the traditional simulation. In case you have replaced a simulational kernel, things are a bit more complicated. As a first stage, you should benchmark your neural network predictions against the outputs of that kernel specifically. We call this *a priori* validation (see Figure 6). Similarity between the predictions here is a minimal requirement and in practice this comparison is relatively easy to do. Thus, it makes sense to also optimize hyperparameters based on such an a priori validation.

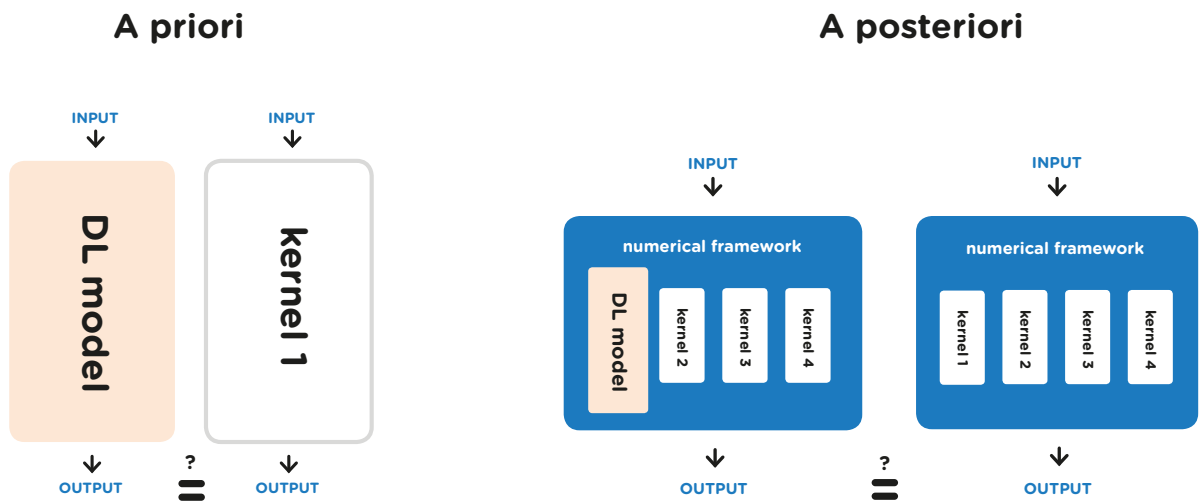

**Figure 6. A priori and a posteriori validation when replacing a single kernel with a deep learning model.**

However, a priori validation is not sufficient. For example, comparing the predictions based on a mean-squared error may show a good similarity between the traditional kernel and neural network predictions. But the distribution of errors may be different, e.g. the neural network may produce *very* wrong results in some rare cases.

The impact of the errors made by a neural network should be assessed in what we call an *a posteriori* validation. For this, the neural network should be integrated into the traditional simulation, replacing the kernel it was trained on. Then, the results of the full simulation (with the incorporated neural network) should be benchmarked against the traditional simulation. Only in this way can we verify that any difference in error distribution does not impact the overall simulation.

# Lessons learned

**Collaboration between AI and domain experts**
One of the key lessons is that collaboration between experts in AI and domain experts is essential at many stages. The domain expert is the key person to identify which aspects of the traditional code provide opportunities for enhancement with deep learning and what properties are known for the inputs and outputs of the model. However, which parts of the code are eligible also depends on how easily the problem can be mapped to an AI solution – which is where the AI expert has the most knowledge.

**Deep learning for HPC: a paradigm shift**
Approaching traditional HPC simulation tasks with deep learning can really be considered a new paradigm. This means it includes the risk of deep-learning based simulations to be rejected by a scientific community. For conservative communities, it might be easier to accept replacement of a small part of a simulation (especially if that part was based on a largely empirical model, as was the case in our turbulence project). On the other hand, replacement of a full simulation has the potential to result in orders of magnitude speedup.

An additional benefit of this new paradigm is that scientific domains that used to employ very different tools and software now suddenly all have very similar tools (such as the popular deep learning frameworks) to solve their problems. This could open up interesting opportunities for collaborations and increase reuse of models/solutions from other scientific domains.

**Good data can be difficult to obtain, even through simulation**
Gathering the data required for a deep learning tasks is one of the key parts in a project and should not be underestimated. Gathering data can take a lot of time, even if it can be obtained through a simulation. This is exemplified by our meteorology project, where obtaining the true unresolved transport (the envisioned output of the neural network) in combination with coarse resolution flow fields required a lot of additional effort. Simply running the model both at low and high resolutions was not an option, since the stochastic nature would have resulted in non-corresponding inputs (the flow fields at coarse resolutions) and outputs (true unresolved transport, derived from high resolution).

**Handling large inputs or large amounts of inputs can be challenging**
Traditional simulations may use fairly large inputs and outputs, such as a 3D wind velocity field in a weather simulation. Generally, that is not a problem in terms of the available memory or amount of disk I/O that needs to be done. Simulations typically only produce one or a few of these outputs per run and correspondingly only have to keep one (or a few) in memory. However, since deep learning is a data driven method, large numbers of samples are needed to train an accurate model. Also, a full batch of samples needs to be able to fit in memory at the same time. Thus, I/O and memory requirements may be very large depending on the type of data used in the model.

Even if inputs and outputs are small, the I/O can be problematic. Parallel file systems such as Lustre are designed for many nodes reading or writing one (or a few files). Thus, they have a substantial I/O bandwidth. However, their capacity to read many small files (and the associated metadata) is not. Thus, if you have a (very) large number of small samples, you may hit file quota limits and/or I/O performance issues. Note that many frameworks have specific file formats that pack numerous (small) samples together to mitigate precisely this issue [11].

**Enhancing HPC simulations with deep learning: a science in itself**
Finally, deep learning for HPC is a very new field of research and since every scientific use case is different, a one-size-fits-all cookbook method to create such enhancements does not exist. This document outlines some general guidelines that may help to structure such a project. In practice, enhancing a traditional HPC code with deep learning is pioneering work and requires a lot of exploration adaptation. Particularly, you may need to go through the stages outlined above iteratively, e.g. exploring multiple ways to map your problem in order to find a solution that works.

# Appendix: SURF deep learning for HPC use cases

## Machine learning for accelerating planetary dynamics in stellar clusters

Maxwell Cai[1,2], Simon Portegies Zwart[1], Valeriu Codreanu[2], Caspar van Leeuwen[2]

[1]Leiden University, [2]SURF

**Introduction**

The evolution of planetary systems in a stellar cluster is largely determined by interactions of these planets with passing stars. These interactions may cause orbits to change, or planets to be ejected from their systems entirely. To understand what planetary systems may be found in the universe, and how the properties of a stellar cluster affect this composition, an accurate simulation of this evolution is required.

Such simulations are however computationally very intensive, since this is a multiscale problem: a stellar cluster can contain thousands of stars, and a multitude of planets. Moreover, changes in planetary orbits occur at much shorter time scales than changes of stellar orbits. To resolve this computational bottleneck, this project investigated the use of neural networks to replace the physics at the smallest scale (namely the interaction of stars with a planetary system), while performing conventional numerical simulation for the large scale (evolution of the stars in the cluster).

**Methods**

In this project, many mappings were explored. The main inputs to the model were the planetary features $a, e$ and $I$, which are the semi-major axis, eccentricity and inclination respectively.

1. **As time series.** The planetary features ($a$, $e$, $I$) were analyzed as a time series using neural networks (e.g. LSTMs). These could then be used to predict future timepoints.
2. **As a pattern recognition problem**. The planetary features ($a$, $e$, $I$) were structured as 2D array. E.g. the first three rows would correspond to the $a$ at time point 1, 2, 3, etc., the 3rd to 6th row would correspond to $e$ at time point 1, 2, 3, etc., and similar for $I$. This 2D array was then analyzed with a convolutional architecture to predict an output vector containing $a$, $e$ and $I$ some 5000 – 10000 years later.
3. **As an image-to-image translation problem.** The planetary feature ($a$, $e$, $I$) were each mapped to their own 2D array, i.e. a 2D array for $a$ would contain e.g. 1 to 10,000 on the first row, 10,001 to 20,000 on the second row, etc. Then, an image-to-image translation network architecture was used to map each of the input arrays to an output array (which would contain the predicted inclinations at the same number of time points in the future). Thus, with this strategy, if one has e.g. 100 Myr of historical time data and one could predict 100 Myr of time points into the future. In principle, the predicted time points could then be used again to make a new prediction (though errors are expected to accumulate with this strategy).
4. **As reinforcement learning (RL) problem.** Here, the planets were interpreted as agents, moving under the influence of their changing environment (a varying gravitational force due to a passing star). When the classical simulation of the star cluster indicated a star to come within a certain threshold distance of a planetary system, the RL model would be invoked to 'decide' how the planetary properties will be changed by this perturber. The *environment* of the planet (i.e. the input of the RL model that determines how the planetary properties will be changed by the encounter) is then determined by features such as the mass, speed and closest point of approach of the perturber, the orientation of its path compared to the plane of the planet, as well as the planetary properties ($a$, $e$, $I$).

**Results**

The aforementioned mappings were explored one by one. Here, we mention the key results

1. Approaching the problem as a time series worked, but it would only enable prediction of future planetary properties up to a few time points in the future. Since planetary dynamics are much faster than stellar dynamics, on the full time scale of the simulation, this didn't help much.
2. Approaching the problem as a pattern recognition problem allows the convolutional network to "see" the evolution history and make longer-term predictions. As long as the ambient environments of the planet remain unchanged in the future, the prediction by the neural network is reliable. However, the ambient environment of a planet is indeed changing slowly (on a typical timescale of a few Myr), this places the limits of the maximum timescale on which this approach can be effective.
3. Approaching the problem as an image-to-image translation allows a neural network to study the underlying physics to some extent. In a typical image-to-image translation problem, an input image is used to condition a generative adversarial network (GAN), which in turn yields a new image accordingly. In this particular problem, the input "image" is the initial condition (e.g. the distance, mass, and speed of the perturber, the semi-major axes, eccentricities, and inclinations of the planets, etc.), and based on these data, a GAN generates the response of the perturbed planetary systems as an "image". This approach, however, fails if the systems are too chaotic. The GAN struggles to handle the sudden changes of planetary orbits because it is a highly nonlinear regime with high dynamic ranges.
4. Approaching the problem as a reinforcement learning (RL) problem allows a RL agent to explore the environment, and therefore understand the physics to a greater extent. The RL agent learns to make predictions on the planetary orbits by trying with different (random) attempts, and subsequently obtain a maximum reward after making a sequence of prediction. The RL is also trained to meet the constraints of relevant laws of physics, such as the conservation of energy and angular momentum. This approach seems works reasonably well in the adabatic regime (i.e. the speed of the perturber is much slower than the orbital speed of the planet). However, the agent is somehow difficult to train.

**Conclusion**

It proved challenging to enhance the simulation of planetary evolution with deep learning. Two properties made this project particularly difficult. First, such systems are chaotic, and therefore their evolutions are very hard to predict. Second, since planets often have multiple encounters, errors made in encounters accumulate.

Many different ways to map the scientific problem were explored in this project, showing the need for exploration and adaptation in *deep learning for HPC* projects. In the end, no satisfying solution was obtained, showing that certain classes of problems may be less suitable for a *deep learning for HPC* approach.

# Machine-learned turbulence in next-generation weather models

Robin Stoffer[1], Chiel van Heerwaarden[1], Damian Podareanu[2], Caspar van Leeuwen[2]

[1]Wageningen University and Research, [2]SURF Open Innovation Lab

**Introduction**

Within the next decade numerical weather prediction models will have a grid spacing fine enough to start resolving the turbulent motions of the near-surface atmosphere [12]. This provides unique opportunities in improving detailed, local forecasts for e.g. the wind speed and air quality. These forecasts are essential for amongst others the wind- and solar energy sector, traffic, (local) governments, and people with respiratory problems.

The next-generation weather prediction models will be based on the large-eddy simulation (LES) technique. With this technique the largest scales of motion of the turbulent flow are being resolved, but in order to take into account the transport at scales smaller than what the model grid can resolve, we rely on a so-called subgrid model. The performance of these models is crucial to correctly represent the lowest 100 m of the atmosphere.

These subgrid models often relate the unresolved fluxes to the velocity gradients in the part of the flow that is resolved. The correlation is often poor which could limit the accuracy of subgrid models. Furthermore, computationally more expensive and advanced subgrid models, such as the scale-dependent dynamical subgrid models, are required to consistently reproduce the near-surface wind and pollutant concentration profiles [13]. The high involved computational cost, however, would make their application in next-generation weather models problematic, and they still rely on assumptions that may introduce biases in the simulation results. Computationally cheap subgrid models, such as the Smagorinsky subgrid model also exist, but these rely on even stronger assumptions and consequently are substantially less accurate. This project aims therefore to use a neural network-based subgrid model, which requires much fewer assumptions than traditional subgrid models. By training neural networks on data generated by computationally costly, but very accurate numerical simulations, our ambition is to obtain a subgrid model that is more accurate than currently used subgrid models, and is (ideally) as fast as Smagorinsky.

**Methods**

Turbulent transport can be considered to be composed of two parts: the part that can be resolved at the scale at which the simulation is run ($\tau_{res}$) and the part that is due to motions that are too small to be resolved at the scale of the simulation ($\tau_{unres}$). The total turbulent transport is then simply given by the sum of the two,

$$\tau_{tot} = \tau_{res} + \tau_{unres}.$$

Note that while the total transport is constant, the part that can be resolved (and the part that remains unresolved) can be seen as a function of the grid spacing at which the simulation is run. Essentially, sub-grid models try to estimate $\tau_{unres}$.

For this project, the well-understood case of turbulent channel flow (with friction Reynolds number $Re_\tau$ = 590) was studied. To generate a training dataset, one might run a simulation at fine scales to determine $\tau_{tot}$ and a second simulation at coarse resolution (that is more realistic for production runs) to provide $\tau_{res}$. $\tau_{unres}$ could then be derived from those two simulations and serve as labels to train a neural network. However, since turbulent flows are stochastic, the $\tau_{tot}$ estimated from fine resolution simulation would correspond to a different *instance* of turbulent flow than the one resolved at coarse resolution.

To resolve this discrepancy, a simulation at high resolution was performed to resolve $\tau_{tot}$ and the same simulation was then downsampled to provide a coarse resolution (corresponding to the same stochastic *instance* of turbulent flow). More specifically: direct numerical simulation (the most accurate numerical method for simulating turbulent flows) with a horizontal grid spacing of ~12.5 m was used to estimate $\tau_{tot}$, which was then downsampled to ~100 m horizontal grid spacing to represent a coarser resolution, LES-based simulation[1]. The procedure is depicted in Figure **7**. All simulations were performed using MicroHH [14].

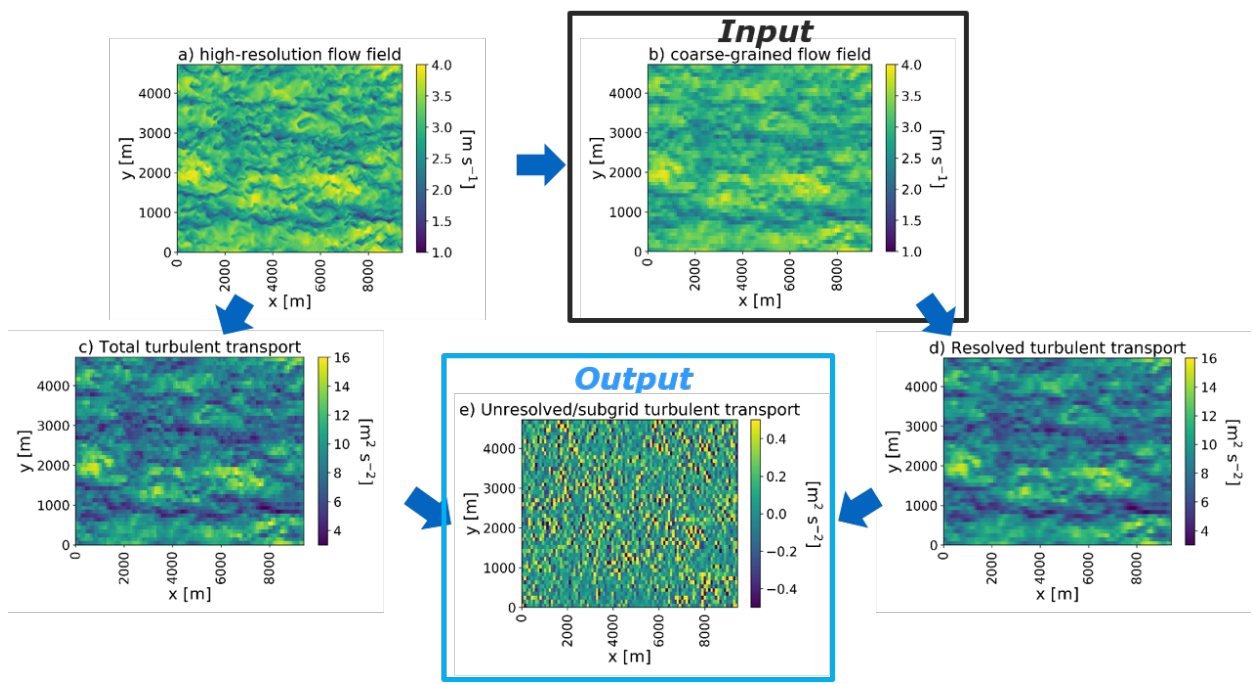

**Figure 7: schematic illustration of how the input and output of the neural network-based subgrid model were obtained from a high-resolution DNS simulation. Shown is a horizontal cross-section of one wind velocity component (u) of the flow field.**

In production, LES-based simulations would use domain decomposition to run on distributed memory systems. To allow for efficient domain decomposition, a hard requirement was imposed on the neural network to predict the unresolved transport based on receptive fields of at most 5 x 5 x 5 grid cells. This is a pretty strong restriction from a deep learning point of view, since a larger field of view might have

---

[1] To be precise, the mentioned grid spacings are based on a rescaling of the original dimensionless grid spacings (≈0.06545δ, where δ is the mid-channel height that is defined to be 1500m). This was done to make the interpretation of the figures easier and more intuitive.

contained more information and allowed more accurate predictions, but this concession was considered essential for the model to integrate well with the traditional simulation. Furthermore, this approach reflects better the typical subgrid modelling approach, where only the locally resolved flow field is considered. A very shallow neural network architecture was adopted in the end to limit the computational complexity. Based on the 5 × 5 × 5 receptive field, a single network initially predicted the 9 components of unresolved transport. However, since the simulation relies on a staggered grid (where the flow variables are not located at the same positions in the grid), the predicted unresolved transports would also be defined on a staggered grid. Eventually, we even switched to predicting two sets of 9 components to prevent that an asymmetric bias was introduced into our results

To validate the outcomes of the neural network, we performed two different tests. First, we performed an offline a priori test, where we simply compared the predicted unresolved transports to the labels for a test set. Secondly, we performed a first online a posteriori test, which aims to assess whether the accuracy of the *overall* model results improve by using a neural-network subgrid model. To this end, we implemented the trained neural network in C++, optimized its inference speed by relying on manual implementation with very efficient matrix-vector libraries (i.e. Intel MKL [15]), and subsequently integrated it within MicroHH.

**Results**

Regarding optimization, we managed to reduce the inference speed of our neural network with about a factor 60x by manually implementing it in C++ with the Intel MKL library, instead of using a frozen Tensorflow inference graph with Python code l. We note though that the neural network is still considerably slower than the Smagorinsky subgrid model. Thus, further optimization has to be performed before it is of practical use.

We also would like to emphasize that this result is not surprising: the neural network contains many matrix-vector multiplications, while the amount of calculations for the Smagorinsky subgrid model is very limited. This illustrates that neural networks can, in some applications, actually increase the total computational cost of the simulation - even if a shallow neural network is used like in our case. It all depends on the computational effort involved in the part of the model code that is being replaced with a neural network. Our case is a good example where the neural network does not actually reduce the total computational cost, but instead can be used as a promising tool to increase the accuracy. To be specific, in our case it allows to learn the unresolved turbulent transport directly from accurate high-resolution simulations, without making stringent prior assumptions that introduce bias.

The a priori validation we did shows well the capability of the neural networks to learn the relationship between the resolved flow fields and the unresolved turbulent transport: we find strong correlations between the unresolved transport predicted by the neural network and the labels. Examples for one component of the unresolved transport at various heights within the channel are shown in Figure **8** and are in the range of 0.7 – 0.8, which is considered exceptionally high for a subgrid model. Note that various instances of high-resolution flow fields could lead to the identical coarse resolution flow fields (it is a many-to-one mapping). Thus, the prediction made by the neural network is a sample (or more likely: an average) of the unresolved transport for each of the potential high-resolution flow fields.

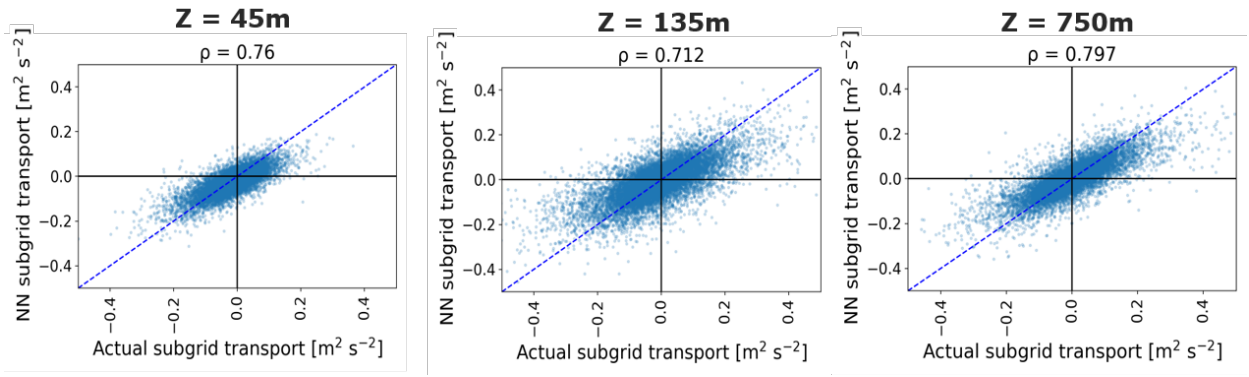

**Figure 8: correlation between labels and neural network predictions for one of the unresolved transport components (i.e. vertical transport of u-momentum), at various heights in the channel.**

A posteriori validation lead to an unstable solution in which the turbulence kinetic energy increases exponentially over time. We are currently investigating the cause of this instability, and potential solutions. A potential source is that the neural network does not dissipate enough energy. If this hypothesis is correct, it may be necessary to enforce more physical constraints during the training of the network.

**Conclusion**
Initial results based on the a priori validation look promising: the neural network-based subgrid model is well able to predict the unresolved turbulent transports. However, a posteriori validation showed that the simulation became unstable when integrating the neural network based subgrid model. This is still a subject of investigation.

It was non-trivial to obtain the data required to train the neural network, since the stochastic nature of turbulence prevented us from simply running independent simulations at a coarse and a fine resolution. The staggered grid added an extra layer of difficulty. Thus, a substantial amount of time was spent on obtaining reliable and suitable training data. This shows the importance of carefully thinking about a strategy to obtain this training data already in the design phase of such a project.

This project also shows the importance of a posteriori validation, when replacing a small part of a simulation. Even if a model works well in an a priori validation, the interaction between the model and traditional simulation may provide additional challenges to overcome.

# Machine-learned radiation in next-generation weather models

Menno Veerman[1], Chiel van Heerwaarden[1], Damian Podareanu[2], Caspar van Leeuwen[2]

[1]Wageningen University and Research, [2]SURF Open Innovation Lab

**Introduction**

Radiative transfer is well-understood as a process, yet its computational burden can take up tens of percents in a typical numerical weather prediction code. Its computation consists of multiple steps. First, the atmospheric properties (gas concentrations, aerosol concentrations, cloud droplets and ice crystals) are converted into optical properties that measure how light behaves inside of a medium. Subsequently the solar and thermal radiation transfer as a function of these properties and the sources (sun light, radiating bodies) is computed.

In this project, we aim to replace the part in which the gases are converted into optical properties by a neural-network-based solver, as this part takes up about 50% of the computational time. The goal of this project is to accelerate that prediction, at the same accuracy as compared to traditional models.

**Methods**

We have generated a wide set of training and testing data based on atmospheric profiles from the Radiative Forcing Model Intercomparison Project (RFMIP [16]) with perturbations around those profiles. To compute the optical properties based on the atmospheric properties, the Rapid Radiative Transfer Model for General circulation model applications - Parallel (RRTMGP [17]) code was used. Since the input and output of the RRTMGP model is identical to that of the envisioned neural network, this provides us with the labeled dataset needed to train the neural network.

Although radiation is essentially a 3-dimensional process, radiative transfer codes (such as RRTMGP) predict the optical properties of each individual grid cell based on only the atmospheric properties of that grid cell at a certain time step. Since we have trained the network to predict – with lower computational costs- the same output as the original radiative transfer code based on the same input, we have also designed the network to predict the optical properties of each grid cell separately. In the atmosphere, of course, there is often a high correlation between the optical properties of adjacent grid cells or of subsequent time steps. Exploiting this correlation might enable us to do more accurate predictions, but may result in a neural network model that does not generalize for arbitrary time steps (i.e. only works for the time steps that are the same as in the training data set). Thus, we decided not to exploit this correlation, so we are able to train a network that is independent of resolution of time step.

We have trained a total of eight different neural networks, since we had to predict four different optical properties and used a separate network for the lower and upper atmosphere. For each optical property, we have to predict 224 or 256 values (depending on the type of radiation) to integrate over the radiative spectrum.

Feedforward multilayer perceptrons are used for the networks, which we built and trained in TensorFlow. A network architecture search was performed, varying the number of layers and neurons per layer. All of these were shallow networks (1 – 3 layers) with the number of neurons per layer ranging between 32 – 128.

For inference, we extracted the weights of the networks from TensorFlow for a manual implementation (based on BLAS) of the feedforward network in the original radiative transfer code. Additionally, we have tried to split up the networks into multiple smaller networks to exploit the correlation within subsets of the 224/256 values. However, this did not outperform our original implementation.

**Results**

Figure 9 shows a few results from the neural network architecture search. While larger networks were clearly more accurate, already the errors from the smallest networks can be considered acceptable: At the surface (≈ 1000 hPa), the average error of RRTMGP compared to a highly accurate line-by-line model is approximately 0.725 W m$^{-2}$ (0.2%) for longwave radiation and 0.026 W m$^{-2}$ (0.01%) for shortwave radiation. The accuracy of the neural networks is thus of the same order of magnitude as the traditional radiative transfer model

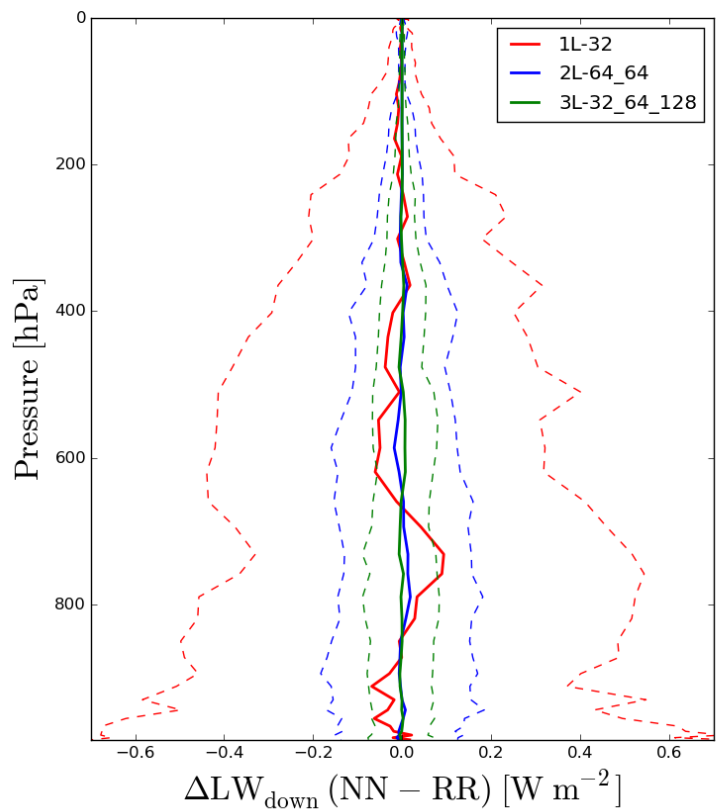

**Figure 9: absolute difference between longwave radiative fluxes based on the neural network-predicted optical properties (NN) and longwave radiative fluxes based on the optical properties of RRTMGP (RR), as a function of atmospheric pressure (which is related to the height in the atmosphere). Shown are the mean and 95% percentile bands for three different network sizes.**

Based on the balance between accuracy and network size, the neural networks with 2 layers of 64 neurons were chosen. Using these networks, our neural network-based solver can reach a speed up of approximately a factor 5 in this calculation, without significant loss in accuracy. As a next step, the neural networks should be validated a posteriori by running a full numerical simulation of the atmospheric with radiative fluxes based on neural network-predicted optical properties.

**Conclusion**

This project showed that even (very) shallow neural networks were already able to predict optical properties based on atmospheric properties, with an accuracy that is on par with traditional models. Moreover, these networks performed predictions approximately 5 times faster than traditional models, providing a fairly substantial speedup.

As a *deep learning for HPC* project, this problem proved relatively straightforward: the inputs and outputs to create a labelled training set could be obtained from the traditional model in a straightforward way. The underlying relationship (i.e. mapping the atmospheric properties to optical properties) may not be trivial to solve analytically, but seems to be reasonably straightforward, considering that shallow networks already provide quite accurate predictions. This allowed us to focus more on optimizing the computational efficiency of the networks, i.e. how to make a network as light as possible while still retaining scientifically acceptable accuracy.

# The Quantum-Event-Creator: generating physics events without an event generator

Sydney Otten[1,2], Sascha Caron[1,3], Wieske de Swart[1], Melissa van Beekveld[1,3], Luc Hendriks[1], Damian Podareanu[4], Caspar van Leeuwen[4], Roberto Ruiz de Austri[5], Rob Verheyen[1]

[1]Institute for Mathematics, Astro- and Particle Physics IMAPP, Radboud University, [2]GRAPPA, University of Amsterdam, [3]Nikhef, Amsterdam, [4]SURF Open Innovation Lab, Amsterdam, [5]Instituto de Fisica Corpuscular, IFIC-UV/CSIC, University of Valencia, Spain

**Introduction**

In the field of high-energy physics and astroparticle physics detectors measure or detect particles. Such particles can be produced in particle collisions created experimentally, such as those at the large hadron collider (LHC) at CERN in Geneva, or naturally by cosmic reactions. A particle physics detector acts like a photo-camera and takes a 'snapshot' of the signals produced by the particles in the detector. Such a picture is called an *event*.

Since the detectors and underlying physics processes are complicated, all comparisons between fundamental theory (such as the Standard Model of particle physics) and measurements are made with the help of so-called event generators. Event generators are programs that generate *artificial events*. These *artificial events* are then compared to real *events* to prove or disprove a theory. For example, in the discovery of the Higgs boson, *artificial events* were generated based on two models: one model that includes the Higgs boson, and one model that excludes the Higgs boson but is otherwise identical. These *artificial events* were then compared to the experimental events measured at the LHC. The measured data showed much better correspondence to the *artificial events* generated with the model that assumed the Higgs boson to exist and thus, the Higgs boson was discovered.

Traditional event generators typically produce events via Monte Carlo simulations, i.e. by random sampling of the underlying probability distributions that govern particle collisions. One of the major issues is that these simulations are computationally very heavy and may take up to 10 minutes for events such as those created at the LHC [18]. Moreover, many such events are needed to obtain enough statistical accuracy to allows comparison against the large amount of experimental data obtained at LHC. Currently, the worldwide LHC computing grid (WLCG) uses approximately 1 million CPU cores and demand is expected to rise with future upgrades of the LHC experiment. A substantial part of the computations performed on the WLCG are event generation.

In this project, we explored replacement of traditional event generators with an event generator that is based on a generative neural network. Event generators based on generative neural networks have the potential to be much faster, and therefore could have an big impact on the compute demands of the LHC experiment.

**Methods**

In this project, three generative tasks were explored

- Simulating 2-particle decay. This toy model has 10 dimensions, corresponding to the energies, momenta and masses of the two particles ($E_1$, $E_2$, $p_{x1}$, $p_{y1}$, $p_{z1}$, $p_{x2}$, $p_{y2}$, $p_{z2}$, $m_1$, $m_2$). A toy Monte Carlo event generator was used to generate the training data.

- Electron-positron collision events in which two leptons (a certain class of particles, see [19]) are produced via an intermediary Z-boson ( [20]). This model has 16 dimensions. A real Monte Carlo event generator was used to generate the training data [21].
- Proton-proton collision events in which a top and anti-top quark are produced ( [22]). This model has 26 dimensions and would represent an event realistic for the collisions in the LHC. In fact, the training used for this task was not generated by us, but it was a subset events from an open dataset containing generated events for the LHC experiment [23].

Two key generative architectures are generative adversarial networks (GANs) and variational autoencoders (VAEs). In this project, several variations of GANs and VAEs were explored. Finally, a novel setup of the variational autoencoder was designed (B-VAE).

The difference between a VAE and B-VAE is somewhat technical. A small hint to those knowledgeable on VAEs is that regular VAEs assume a normally distributed prior for the latent space. If the encoding from the observations in the end is not normally distributed, any decoder based on the same assumption will fail to generate realistic samples. Thus, the B-VAE *aims* for a normally distributed prior, but at the same time *stores* the deviation from a normal distribution. In such a way a non-normal prior can be assumed for the generation task that proves to generate much more realistic events. The full details can be found in [18].

Though many variations of GANs and VAEs were explored, the key results presented here correspond to the following combinations of generative architectures and generational tasks:

- A traditional GAN, traditional VAE and the B-VAE were used to simulate 2-particle decay.
- A traditional VAE and the B-VAE were used to simulate the electron-positron collision events producing two leptons via an intermediary Z-boson.
- The B-VAE on the proton-proton collision events that produced a top and anti-top quark.

One of the things that was non-trivial in this project is to assess the performance of these generative networks. There are several metrics for generative networks, but they are often targeted at image generation. Thus, performance was assessed by defining metrics relevant to the scientific task, namely generating realistic physics. The main way to check this was to verify that the generated particles follow the same distributions as the training set.

For example, for the two particle decay, we compare histograms of the physical parameters (momenta, energy, etc.) of the events generated by neural networks to those generated by a conventional Monte Carlo event generator. Furthermore, for the electron-positron collisions producing two leptons, we also check the correlation between the angles of the produced leptons. For the proton-proton to top anti-top generation task, distributions of more complicated physics parameters were validated (these are usually composite or derived parameters and more commonly used in real experiments such as the LHC).

Finally, since the neural network architectures used in this research have a large number of degrees of freedom, it was important to check that these networks were actually learning *physics* and did not just learn to *represent* (only) the events on which they were trained (in the latter case, generalizability would be very limited). To do so, the variational autoencoder was also tasked with reconstructing random inputs. If random inputs are more poorly reconstructed than true inputs, it shows that the autoencoder architecture is actually learning the structure in the real inputs.

**Results**

Simulation times were $O(10^8)$ faster for the neural network based simulations. Figure **10** demonstrates that the B-VAE could generate all types of events with an accuracy that is close to that of to the traditional Monte Carlo simulations. Furthermore, Figure **11** shows that the network can learn physical laws such as the conservation of momentum, without these explicitly being incorporated in the network. Finally, Figure **12** demonstrates that the B-VAE does not simply learn to represent *only* the input samples, but learns the structure in particle physics events and thus has generalizing potential.

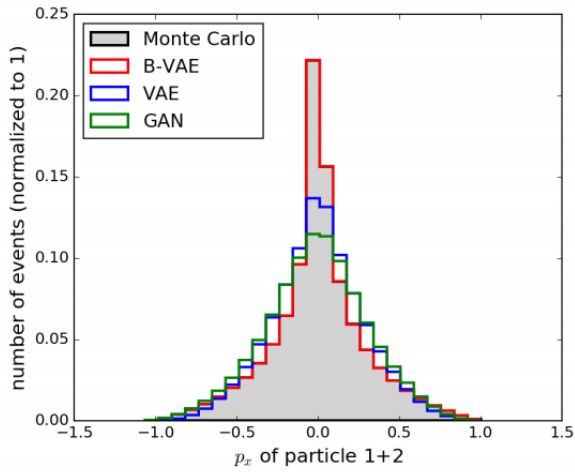
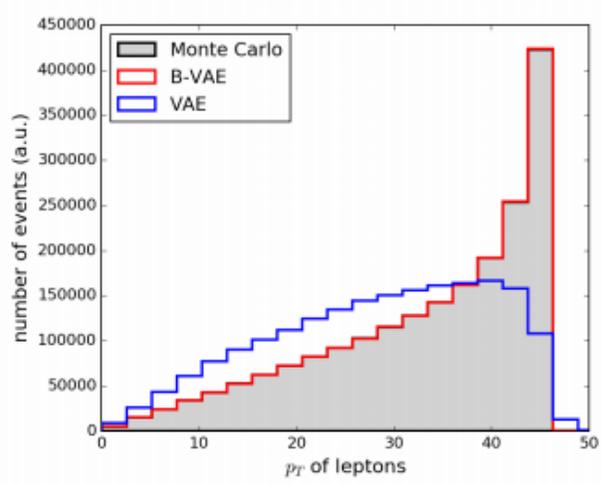
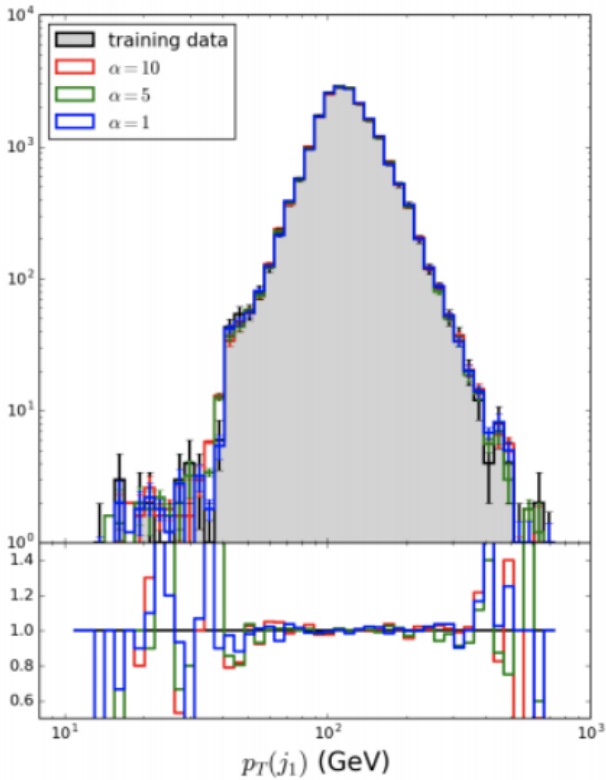

Figure 10: momentum distributions for traditional Monte Carlo simulations compared to various neural network based generators for two-particle decay (top left), electron-positron collision producing two leptons (top right) proton-proton collisions producing a top and anti-top quark (bottom left).

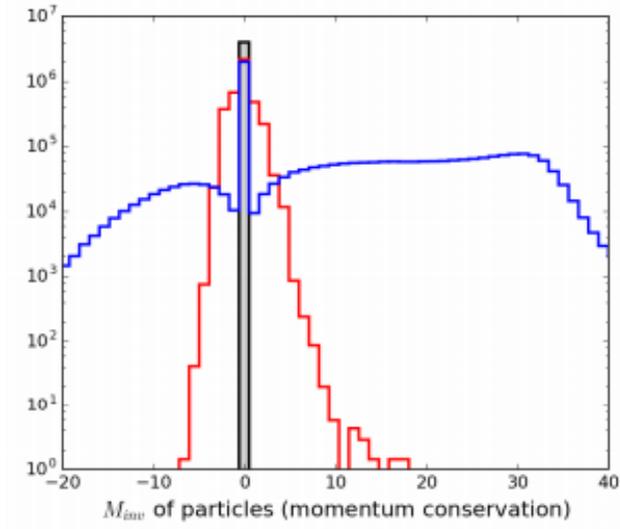

Figure 11: distribution of $M_{inv}$ (the invariant mass) of the particles in the generated for electron-positron collisions producing two leptons. Note that an $M_{inv} = 0$ represents momentum conservation.

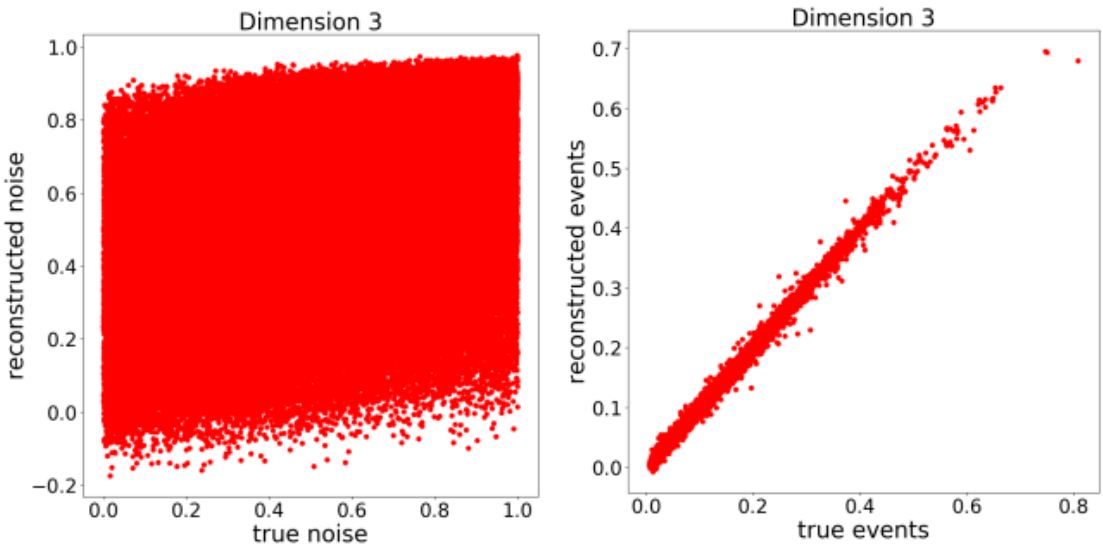

Figure 12: Input compared to reconstruction for uniform noise $x \sim U(0, 1)$ (left panel) and for real events (right panel). Perfect reconstruction would correspond to an identity function.

**Conclusions**

This research has shown that groundbreaking speedups in simulation time are possible when using generative neural networks for particle event generation tasks. Additionally, these networks are able to represent the underlying physics to a similar degree as traditional simulators.

Furthermore, this research shows that traditional validation metrics may not be as meaningful for *deep learning for HPC* approaches. Those properties important for particle physicists were carefully considered and validated. Furthermore, conservation laws known from fundamental physics allowed for an additional sanity check that the neural network is learning physics.

# 3DeepFace: Distinguishing biological interfaces from crystal artifacts in biomolecular complexes using deep learning

Cunliang Geng[1], Li Xue[1], Francesco Ambrosetti[1], Alexandre Bonvin[1], Valeriu Codreanu[2], Caspar van Leeuwen[2]

[1]Utrecht University, [2]SURF Open Innovation Lab

**Introduction**

Many essential cellular functions are mediated through specific protein-protein interactions. The physiological functions associated with those interactions are closely related to their three-dimensional (3D) structure. Hence, knowledge of the biologically-relevant 3D structure of a protein-protein complex is relevant in order to understand its function. Currently, 3D structures of proteins and their complexes can be experimentally determined through several techniques, with the most common one being X-ray crystallography. In this technique, proteins are first crystallized and then exposed to X-rays to get electronic densities, in which atoms can be built to represent the 3D structure. However, in the crystal, along with the biologically relevant interfaces, crystallographic ones are also created. These correspond to non-specific interactions and are artifacts of the crystallization process.

The identification of the biologically relevant interface is therefore of paramount importance and usually non trivial due to the high complexity of biomolecular interactions. The problem is schematized in Figure **13**.

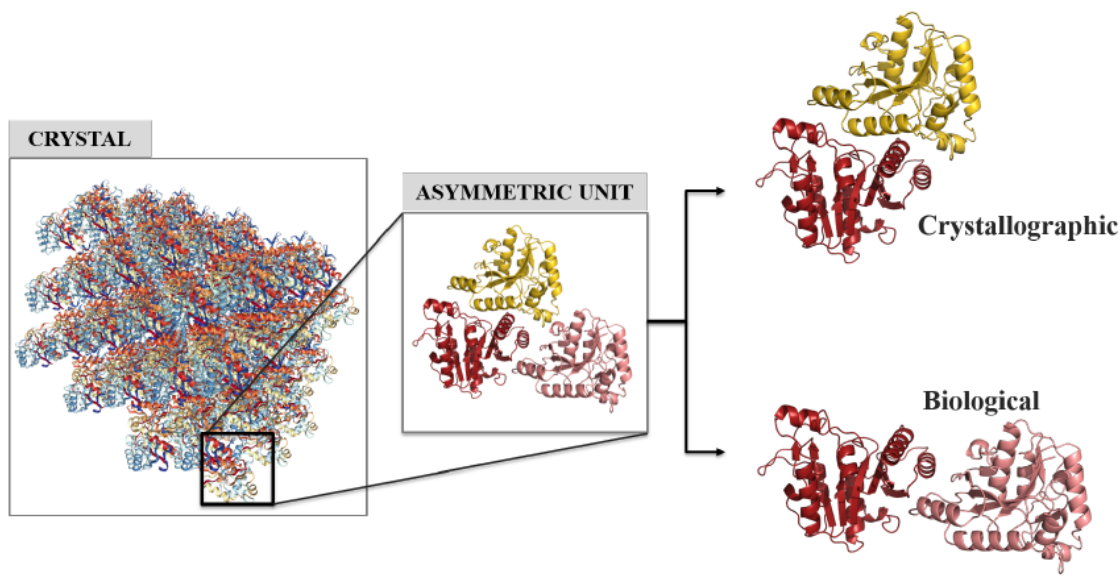

**Figure 13. Illustration of crystallographic and biological interfaces present when proteins are studied in crystallized form. It is a-priori unknown which interfaces in the crystal unit are crystallographic artifacts and which are biological.**

To tackle this problem, traditional wet-lab experiments are often required but they are usually time-consuming and laborious. Computational approaches offer a faster and cheaper alternative solution.

Over the years, several computational methods have been developed to distinguish biological interfaces from crystallographic ones, leveraging on a variety of geometrical, energetic and evolutionary properties of biomolecular interactions and relying on both machine learning and non machine learning approaches. Among those, EPPIC [24], PISA [25] and PRODIGY-CRYSTAL [26] [27] show the highest accuracy. EPPIC is based on evolutionary and geometrical information, PISA estimates the binding energy between the two molecules, while PRODIGY-CRYSTAL, the only machine learning based method among them, uses a combination of structural (residue-residue contacts, buried surface area and non-interacting surface features) and energetic properties (Electrostatic, Van der Waal and Desolvation). Among them PRODIGY-CRYSTAL shows the highest accuracy demonstrating the great potential of machine learning approaches in classifying crystallographic versus biological interfaces.

Here we aim at improving the classification accuracy by representing the protein interfaces as 3D images and by developing a deep learning model to distinguish between crystallographic and biological interactions.

**Methods**
As already mentioned, several features have been previously identified as good discriminants between biological and crystallographic interfaces with one of the most interesting ones being the evolutionary conservation.

In our method we treat the problem as an image classification problem by representing the interface between proteins as a 3D image. Since Interfaces of different protein-protein complexes have different shape and size, they cannot be used directly as input of 3D CNN. To solve this problem, for each protein-protein complex we place a fixed-size 3D grid box around the interface between the two proteins. Each interface is therefore represented as a 3D grid of points. The evolutionary conservation of each residue encoded in the form of position-specific-scoring matrix (PSSM) [28] was then calculated and mapped on the grid points.

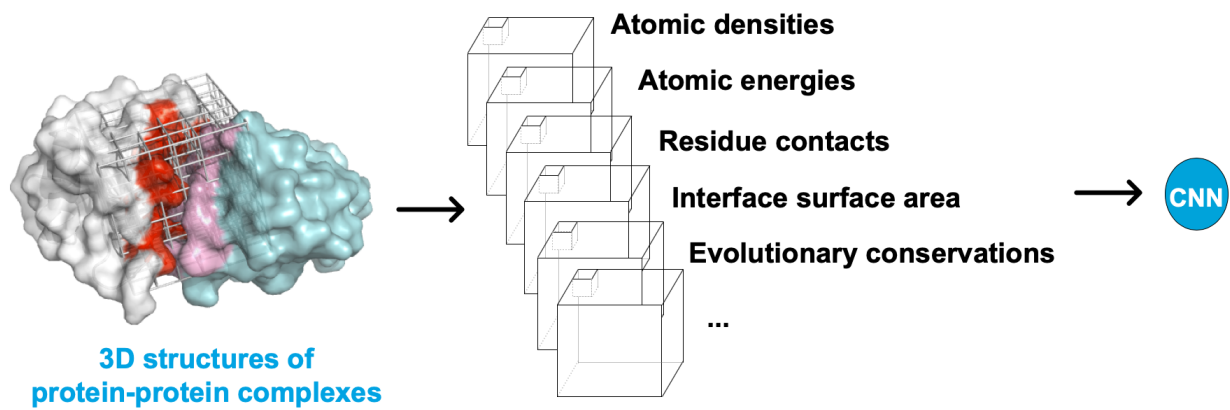

Figure 14. Interfaces can be mapped onto 3D grids, transforming 3D atomic coordinates into 3D images that can be used in image classification enhanced

**by CNN. A 3D convolutional network architecture was used here for the classification task. While not particularly deep (the network contained two convolutional, and one fully connected layer), the computational complexity added by the third dimension mean training times can be substantial.**

Two datasets have been used to train and test our model. For training we used the MANY dataset [29] which consist of 2831 biological complexes and 2913 crystallographic ones while for testing the DC dataset [30] has been used. It contains 80 biological complexes and 80 crystallographic ones. Moreover, in order to enhance the predictive power of our method each structure of the MANY dataset has been randomly rotated 10 times. Since they both datasets (MANY and DC) are balanced we used the accuracy as parameter to assess the performance of our model.

**Results**

Figure **15** shows the results in terms of accuracy of the CNN predictor as a function of the number of Epochs on the training and test sets. The training set was split into training (80%) and validation sets (20%). From Using early stopping as regularization technique at epoch 3 in order to avoid overfitting, our model achieves an accuracy of 83% on the test set outperforming both PRODIGY-CRYSTAL and PISA which have been demonstrated to be the best performing methods on this dataset with 74% and 79% accuracy, respectively [31].

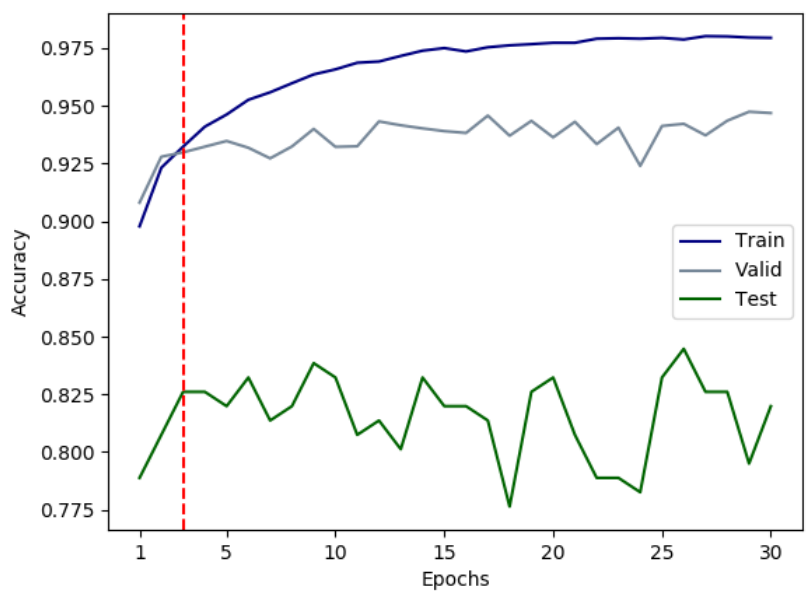

**Figure 15. Performance (accuracy) of 3DeepFace in classifying biological versus crystallographic interfaces as a function of the number of epochs. The accuracy for the training, validation and test sets are reported in blue, gray and green, respectively. The red line indicates the epoch 3.**

**Conclusions**

We have presented here a novel method to represent protein interfaces as a 3D images and demonstrated that those can be used to train a CNN classifier for distinguishing biological from crystallographic interfaces. 3DeepFace on the test set outperforms the current best methods, demonstrating the potential of machine learning and HPC to address biological problems. Further progress might be achieved by further optimizing both the network architecture and the hyperparameter selection along with the use of additional features for the classification.

**Acknowledgements**

We want to thank Erik van der Spek for editorial work and Sanne Koenen for creating various figures and taking care of overall layout.